\documentclass{ieeeaccess}
\usepackage{cite}
\usepackage{amsmath,amssymb,amsfonts}
\usepackage{gensymb}
\usepackage{algorithm}
\usepackage{algpseudocode}
\usepackage{listings}
\usepackage{graphicx}
\usepackage{textcomp}
\usepackage{comment}
\usepackage{colortbl}
\usepackage{color}
\usepackage{soul}
\definecolor{gray}{rgb}{0.3,0.3,0.3}
\definecolor{gray_2}{rgb}{0.8,0.8,0.8}

\def\BibTeX{{\rm B\kern-.05em{\sc i\kern-.025em b}\kern-.08em
    T\kern-.1667em\lower.7ex\hbox{E}\kern-.125emX}}
\begin{document}
\history{Date of publication xxxx 00, 0000, date of current version xxxx 00, 0000.}
\doi{10.1109/ACCESS.2017.DOI}

\title{Cyber-Security Internals of a Skoda Octavia vRS: A Hands on Approach}
\author{\uppercase{Colin Urquhart}\authorrefmark{1}, 
\uppercase{Xavier Bellekens}\authorrefmark{1}, \IEEEmembership{Member, IEEE},
\uppercase{Christos Tachtatzis}\authorrefmark{2},\IEEEmembership{Senior Member, IEEE},
\uppercase{Robert Atkinson\authorrefmark{2},\IEEEmembership{Senior Member, IEEE},
\uppercase{Hanan Hindy}\authorrefmark{1},\IEEEmembership{Member, IEEE}
and Amar Seeam}.\authorrefmark{3},\IEEEmembership{Member, IEEE}}
\address[1]{Abertay University, Division of Cyber-Security, Dundee, Scotland}
\address[2]{University of Strathclyde, Dept. Electronic and Electrical Engineering, Glasgow, Scotland}
\address[3]{Middlesex University, Dept. Computer Science, Uniciti, Flic-en-Flac, Mauritius}

\markboth
{Urquhart \headeretal: IEEE ACCESS}
{Urquhart \headeretal: IEEE ACCESS}

\corresp{Corresponding author: Xavier Bellekens (e-mail: xavier.bellekens@ieee.org)}

\begingroup
\newlength{\xfigwd}
\setlength{\xfigwd}{\columnwidth}
\endgroup

\begin{abstract}
The convergence of information technology and vehicular technologies are a growing paradigm, allowing information to be sent by and to vehicles. This information can further be processed by the Electronic Control Unit (ECU) and the Controller Area Network (CAN) for in-vehicle communications or through a mobile phone or server for out-vehicle communication. Information sent by or to the vehicle can be life-critical (e.g. breaking, acceleration, cruise control, emergency communication, etc\ldots). As vehicular technology advances, in-vehicle networks are connected to external networks through 3 and 4G mobile networks, enabling manufacturer and customer monitoring of different aspects of the car. While these services provide valuable information, they also increase the attack surface of the vehicle, and can enable long and short range attacks. In this manuscript, we evaluate the security of the 2017 Skoda Octavia vRS 4x4. Both physical and remote attacks are considered, the key fob rolling code is successfully compromised, privacy attacks are demonstrated through the infotainment system, the Volkswagen Transport Protocol 2.0 is reverse engineered. Additionally, in-car attacks are highlighted and described, providing an overlook of potentially deadly threats by modifying ECU parameters and components enabling digital forensics investigation are identified. 

\end{abstract}


\titlepgskip=-15pt

\maketitle

\section{Introduction}
\label{sec:introduction}
\PARstart{T}{he} average family car produced between the 1960's - 1980's had limited electronics and technological systems installed. As technology advanced, the additions to the ECU, remote central locking, CAN bus network and electronic safety systems became standard~\cite{van2011canauth}, however, the speed of technological advancement through mobile applications and increased communication services, to and from car companies, has resulted in an increased attack surface and security vulnerabilities. This is due to the lack of security consideration for vehicular technologies and associated connected systems~\cite{8666649}.

\begin{figure*}[!th]
	\centering
	\includegraphics[width=1.0\linewidth]{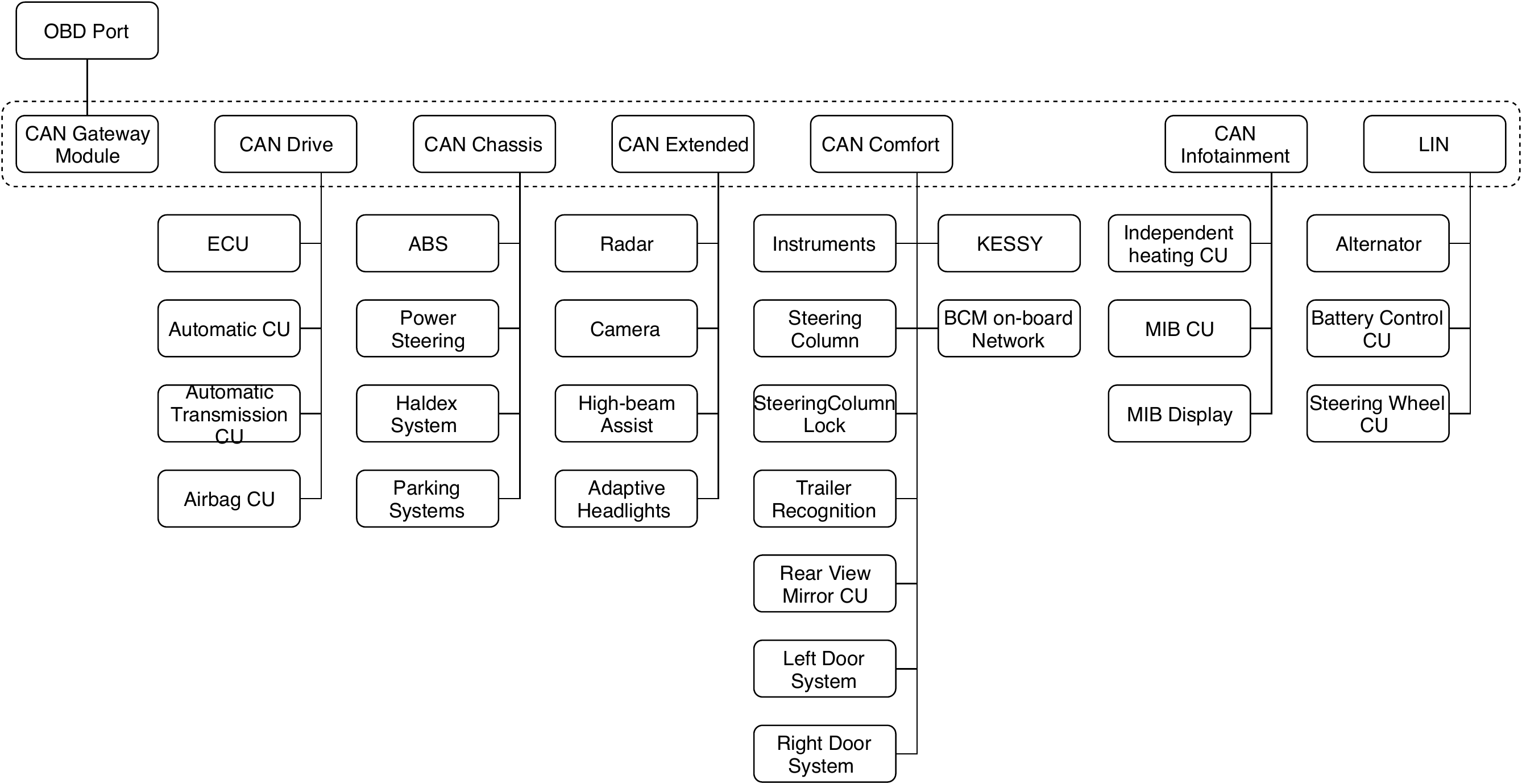}
	\caption{Skoda CANbus Overview}
	\label{fig:canbus-overview}
\end{figure*}

In 2015, Samy Kamkar~\cite{osborne_2015}~\cite{qin2016intrusion} presented \textit{OwnStar} an off-the-shelf device, able to jam, store and replay key fob signals of General Motors vehicle, hence, bypassing the rolling code transmission. General Motors released an update to avoid those types of attack later-on that year. In 2016, Keen Security Lab successfully compromised a Tesla~Model~S~\cite{nie2017free}. The attack was achieved remotely and provided control over the car, both in parking and driving modes. The car was unmodified and running the latest firmware at the time. In 2017, Keen Security Lab successfully found numerous 0-day attacks in the Tesla~Model~X~\cite{blog_2019}.
\newline \newline
While researchers have explored the security of vehicular technology at various level, numerous car manufacturers use legacy systems and are unable to update the security of cars currently in circulation, nor are they able to control the different attack vectors on cars running legacy systems (e.g. infection through a rogue garage)~\cite{bellekens2015cyber}. 
\newline \newline
With over 1.25 million Skoda Vehicles sold in 2018~\cite{Skoda_2019}, this paper investigates legacy systems of the Skoda Octavia vRS 2017, while providing a global outlook on cyber-security vulnerabilities and digital forensic elements of current vehicular technologies. Exploring legacy vehicular technologies is critical, as a typical usage time frame for a car is of 5-10 years~\cite{youtube}, while the development time frame is around 3 to 6 years~\cite{youtube}. Moreover, vehicular manufacturer often operates around the world, hence, delivering different system to different continents based on user demand and requirements, i.e. pushing different software updates and rollouts (e.g. infotainment systems may be different in China and the EU)~\cite{youtube}. Finally, current code in vehicles accounts for over 10.000.000 lines of code, including 3rd party supplier code, increasing the attack surface~\cite{youtube}.

The contributions of this work are four fold:
\begin{itemize}
    \item A Vehicular Security Testing Methodology is presented; enabling to identify key components to test under various circumstances.
    \item Vulnerabilities in Skoda Octavia vRS 2017 are explored.
    \item A Proof of concept algorithm reversing the VW TP 2.0 protocol to access and modify the ECU is provided, further highlighting the danger of off-the-shelf remapping devices.
    \item Components able to yield vehicular forensics information are identified and presented.
\end{itemize}

The remainder of this paper is organised as follows; Section~\ref{sec:relatedwork} provides an overview of the related work. Section~\ref{sec:investigationprocess} highlights the investigation process followed in this research, as well as, identifies a methodology for security testing. Section~\ref{sec:key} demonstrates security vulnerabilities of the current rolling code car fobs used by the vehicle. Furthermore, Section~\ref{sec:info} highlights vulnerabilities in the infotainment systems. Section~\ref{sec:obd} explores the OBD-II port to modify key parameters of the vehicle. Section~\ref{sec:ECU} demonstrates the modification of the ECU through the OBD-II port. Moreover, Section~\ref{sec:forensics} provides an overview of useful components for digital investigators. Finally, the paper concludes with Section~\ref{sec:conclusion}.

\section{Related Work}
\label{sec:relatedwork}
Over the last couple of years, vehicular technologies have been at the centre of attention with autonomous cars fully embodying the Internet of Things (IoT) concepts, enabling remote control of certain aspects of the cars such as heating, the state of brake pads, remote locking, etc\ldots. These features have drastically impacted the attack surface of modern cars. Leading to numerous flaws being uncovered over the years. In 2013 Valasek~\textit{et al.}~\cite{193260} published a manuscript detailing vulnerabilities in Volkswagen vehicle immobilisers. The paper discussed the reverse engineering of the device and demonstrated that vehicles could be stolen without much effort. 

In 2014 Miller~and~Valasek~\cite{miller2013adventures} published the ``Adventures in Automotive Network and Control Units'' confirming the possibility for a remote attacker to execute code on a vehicle Engine Control Unit (ECU). The authors detailed potential modifications an attacker would be able to carry out remotely on a Ford and Toyota test vehicle. The manuscript further highlights critical parameters they were able to modify while the car was in motion. These included, changing the steering during auto park on the Ford and fooling the pre-collision system into slowing down or stopping the car whilst in motion on the Toyota.

In~\cite{smith2016car}, the authors highlight the various vulnerabilities within vehicles as well the cars various systems including reverse engineering the Controller Area Network (CAN) bus and ECU. The authors further highlight the possibility of threats surrounding Vehicle-to-Vehicle (V2V) communication and signalling. 

In~\cite{garcia2016lock}, an overview of the security vulnerabilities with modern key-less entry systems is provided. The researchers successfully managed to gain access to a number of test vehicles from the VW group. The paper detailed an attack on the Hitag2 cryptographic key and provide a discussion on cloning the remote control, resulting in the authors gaining access to the vehicles.

\begin{figure}[t]
	\centering
	\includegraphics[width=0.5\linewidth]{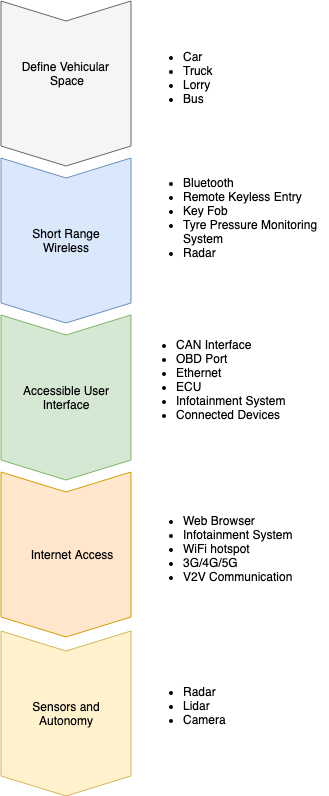}
	\caption{Vehicular Security Testing Methodology}
	\label{fig:sectTesting}
\end{figure}

\cite{connected18} focuses on gaining remote access to the Volkswagen Golf GTE and an Audi A3 e-tron vehicle. The authors demonstrate some methods and tools used to gain access, however, due to responsible/ethical disclosure, they do not disclose all methods used to gain access to the vehicles. When discovered, the vulnerabilities were raised to Volkswagen which in turn reviewed the research paper and advised that all the identified security vulnerabilities be subsequently resolved in an update to the infotainment system.

As described, the attack surface in vehicular technologies increases apace with the number of sensors and actuators deployed. The next section provides an overview of the sensors and actuators in the test vehicle presented in this manuscript, furthermore, a methodology for testing the cyber-security of vehicular technologies is presented alongside components able to yield data in a digital forensic investigation based on our cyber-security findings.

\section{Investigation Process}
\label{sec:investigationprocess}
In this section, the vulnerabilities of the Skoda Octavia vRS are highlighted. This manuscript focuses on end-user accessible interfaces, such as the key-fob, the infotainment system, the On Board Diagnostic~(OBD) interface and the ECU as shown in Figure~\ref{fig:canbus-overview}. These interfaces have been chosen, as they are accessible by both end-users and mechanics, furthermore, these interfaces can also be linked to connected off-the-shelf devices for fault diagnosis. While this manuscript aims at assessing vulnerabilities of end-user accessible interfaces, we provide a generalised cyber-security testing model fully transferable to other vehicular technologies. 

\subsection{Cyber-Security Testing Model}
While vehicular technologies have specificities, they are also composed of similar components. Generally, modern cars include at least one short-range wireless communication, provide the users with accessible interfaces such as OBD2, the CAN interface or the ability to connect a phone to the car through a USB. Furthermore, numerous cars have some form of connectivity enabled, such as tracking through 3G networks or have a browser integrated to the infotainment systems. Furthermore, all cars have an ECU and feature sensors such as oil gauge, breaking pads wear sensors, etc. All these characteristics are presented within the methodology in Figure~\ref{fig:sectTesting}. The method aims at identifying the attack surface. Step I) \textit{Define Vehicular Space:} identifies the vehicle based on the context space; as an example, a refrigerator truck will have different sensors than a garbage truck. Step II) \textit{Short Range Wireless:} defines the attack surface based on all sensors using short-range wireless technologies such as 433MHZ, Bluetooth, etc. Step III) \textit{Accessible User Interface:} enables to list all the accessible user interfaces, ranging from accessible ports on the car as well as the infotainment system and USB ports. Step IV) \textit{Internet Access:} allows to identify all connection points to the internet within the vehicle, including the software that allows to browse the internet. Step V) \textit{Sensors and Autonomy:} provides an opportunity to list Lidars and other sensors the car uses to drive and autonomously navigate areas. 

This methodology was used to identify the attack surface for the Skoda Octavia vRS.  In this manuscript, the short-range wireless system was considered, more specifically the key fob. Different user interfaces were also considered, such as the OBD II port, the connected devices, the ECU and the Infotainment systems. Finally, the Internet Access was tested through the Infotainment system and the Web browser. The Skoda Octavia did not include any Radar or Lidar, however, different sensors were altered through the User Interface (OBD II Port). 

\section{Key Fob Examination}
\label{sec:key}
\begin{figure}[t]
	\centering
	\includegraphics[width=1.0\linewidth]{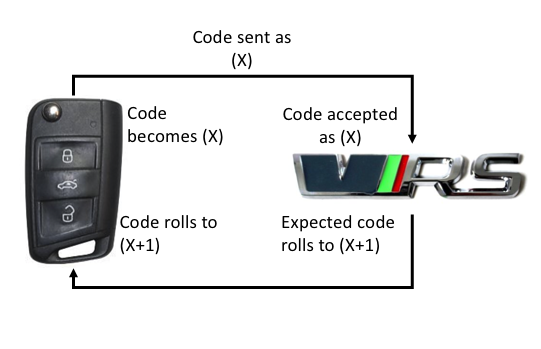}
	\caption{Rolling Code Overview}
	\label{fig:rc}
\end{figure}

One of the most integral aspect to any car has always been the key fob, yet the method for unlocking vehicles has largely remained unchanged since the development of key-less entry, with the exception of moving from a ``fixed" code to a ``rolling" code. Fixed code, re-use the same code over the longevity of the key, whereas rolling code, changes the code each time the key and the car communicate. Figure~\ref{fig:rc} illustrates the sequence of a rolling code transmission.

In this section, the feasibility of a replay attack against rolling code enabled key fob is investigated. This attack when executed correctly is highly effective and has been demonstrated by thieves on surveillance camera~\cite{stealing_2018}, however, to the best knowledge of the authors, a hands-on approach has never been documented. This section demonstrates this attack in details and is achieved in three distinct steps; Step I) A noise/jamming signal with the aim of blocking communications between the car and the key when the legitimate user presses a button (e.g. open car, close car) is created. Step II) A Software Defined Radio (SDR) peripheral is used to record the communication between the key fob and the car, including the rolling code. Finally, Step III) The device replays the rolling code obtained from the key fob to the car.

\begin{figure}[b]
	\centering
	\includegraphics[width=1.0\linewidth]{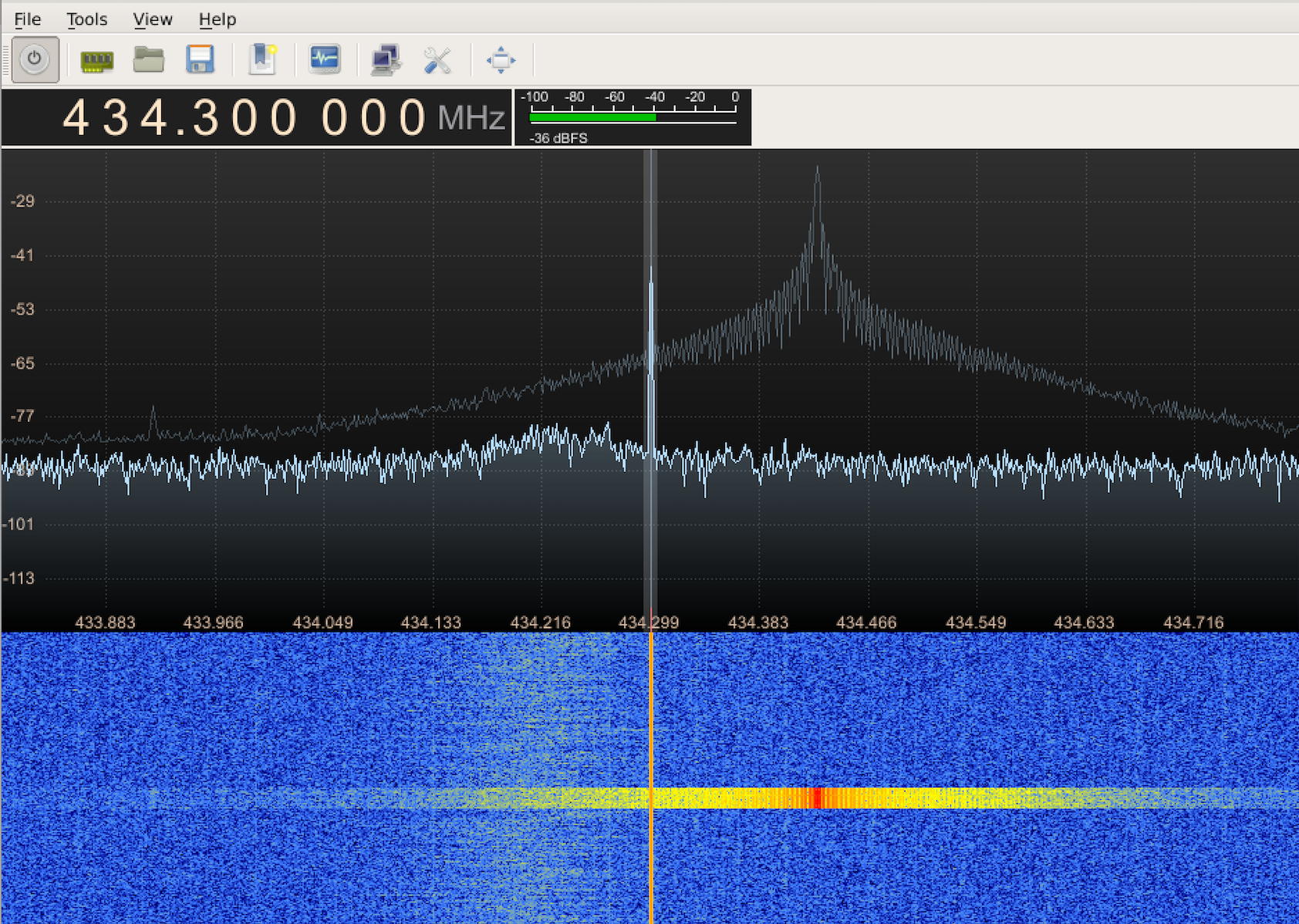}
	\caption{GQRX Frequency Confirmation}
	\label{fig:screen-shot-2018-06-06-at-14}
\end{figure}

In the United Kingdom, keys operate at a frequency of 434MH, to this end the SDR receiver ix tuned to 434.3MHZ. Figure~\ref{fig:screen-shot-2018-06-06-at-14} shows that transmissions between the key fob and the car are recorded between 434.383MHz and 434.466MHz.

\begin{figure}[t]
	\centering
	\includegraphics[width=1\linewidth]{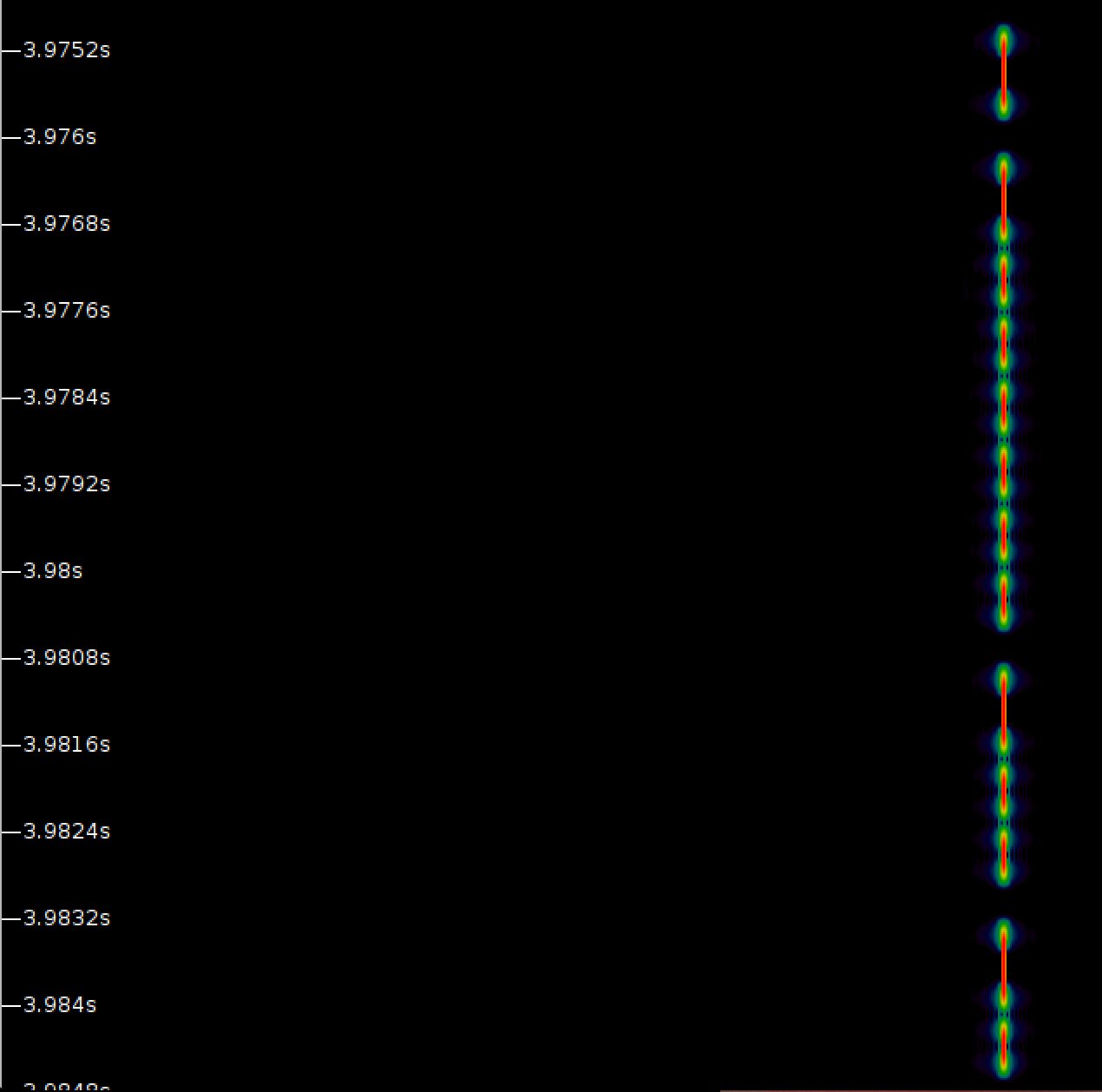}
	\caption{Key Fob Signal Modulation}
	\label{fig:inspectrum}
\end{figure}

Following the tuning of the SDR receiver, the signal between the key fob and the car is captured. Figure~\ref{fig:inspectrum} shows the modulation of the signal.

The signal in Figure~\ref{fig:inspectrum} is shown to incorporate an Amplitude Shift Keying (ASK) for the transmission, specifically On-Off-Keying (OOK). This form of modulation works as follows; each ``on" pulse is categorised as a "1" while each ``off" pulse represents a "0". In order to fully capture the key fob signal, it is necessary to introduce noise, in order to prevent the car from accurately recognising the transmission. This is achieved by developing an algorithm for the YARDstick One. The YARDstick One, is a low speed USB transceiver, which enables to capture and record transmission under 1GHz. 

\begin{algorithm}[b]
\caption{}
\label{alg:skoda}
\begin{algorithmic}[1]
\State $x$ = ``define frequency"
\State $w$ = ``define channel width"
\State $l$ = ``define packet length"
\item[]
\State set frequency to $x$
\State set width to $w$
\item[]
\State set modulation
\State set modem rate
\State set the data which is sent on the channel
\State set packet length to $l$
\State set max TX power
\State start transmitting
\State take input to stop the signal
\State print ``completed"
\end{algorithmic}
\end{algorithm}

Algorithm~\ref{alg:skoda} displays the pseudocode to create a wide channel of noise. This enables to generate a transmission of sufficient width at a frequency of 434MH, forbidding the receiver to accept the transmission. Furthermore, this crucial step ensures that the key fob is not de-authenticated during the process.

\begin{figure}[tb]
	\centering
	\includegraphics[width=1\linewidth]{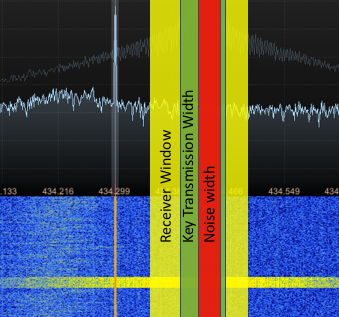}
	\caption{Noise Goal}
	\label{fig:noise-goal}
\end{figure}

Figure~\ref{fig:noise-goal} shows the noise created. As seen, the noise band is large enough to cover the transmission width of the key fob. Whilst the noise is being transmitted, GNU Radio is used to capture the Raw IQ data from the transmission, filter the noise signal and replay the transmission to the car.

\begin{figure}[b]
	\centering
	\includegraphics[width=1\linewidth]{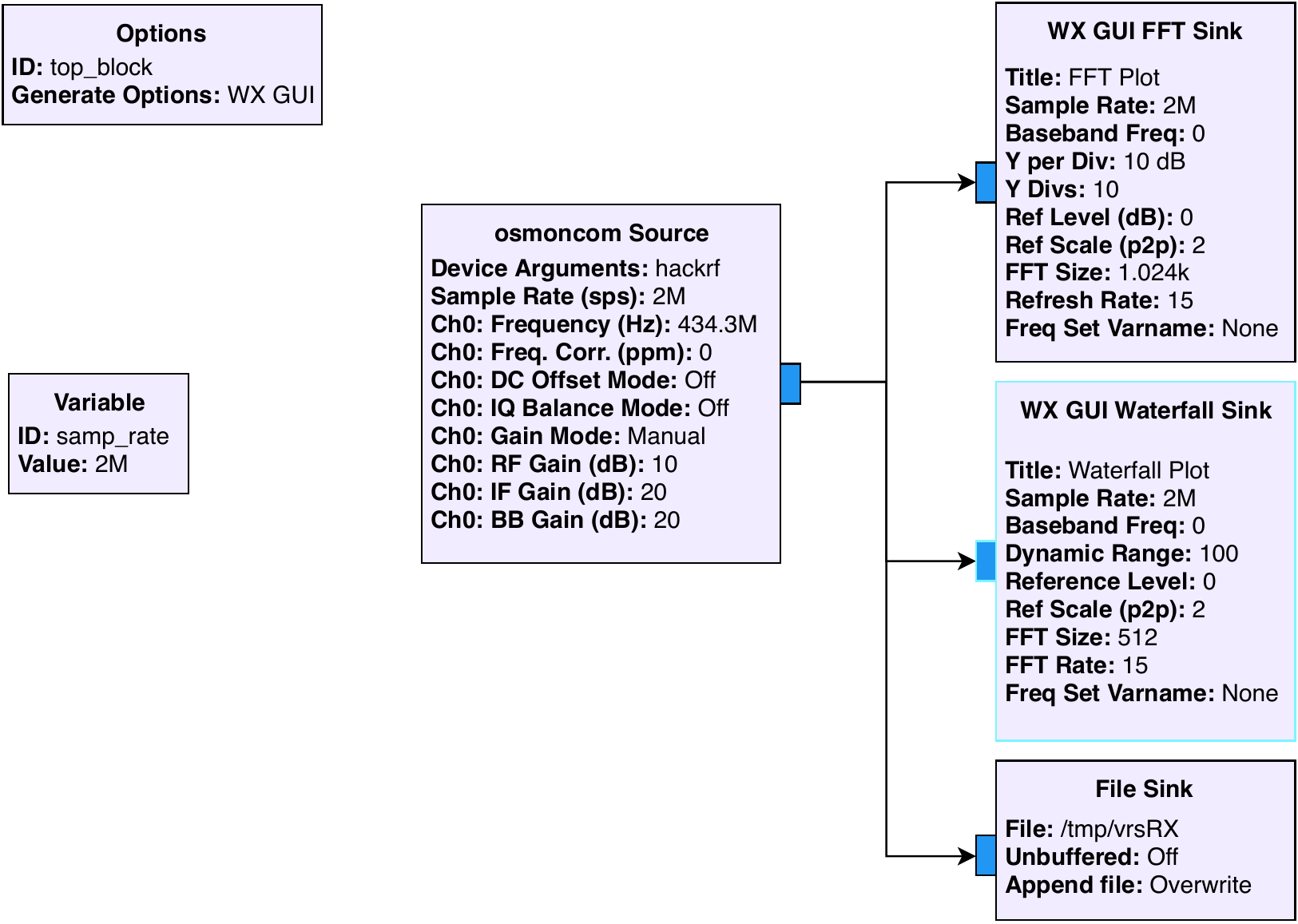}
	\caption{vRSRX}
	\label{fig:vrsrx}
\end{figure}

Figure~\ref{fig:vrsrx} shows GNU Radio configured with a sample rate of $2e^6$, with the frequency set for 434.3m Hz. This configuration, enables both to capture the key transmission, but also to generate the background noise as well as re-transmit the signal from the key fob afterwards. 

\begin{figure}[t]
	\centering
	\includegraphics[width=\linewidth]{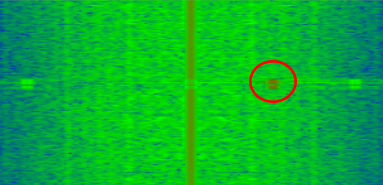}
	\caption{Signal and noise capture}
	\label{fig:signalandnoise}
\end{figure}

It is important, however, to note that the signal captured contains the background noise generated, hence it is critical, to filter the signal captured. As shown in Figure~\ref{fig:signalandnoise} the car would be unable to recognise the valid transmission due to the background noise. It is, therefore, necessary to filter the noise down prior to replaying the key fob transmission as highlighted in red in Figure~\ref{fig:signalandnoise}.

\begin{figure}[b]
	\centering
	\includegraphics[width=1\linewidth]{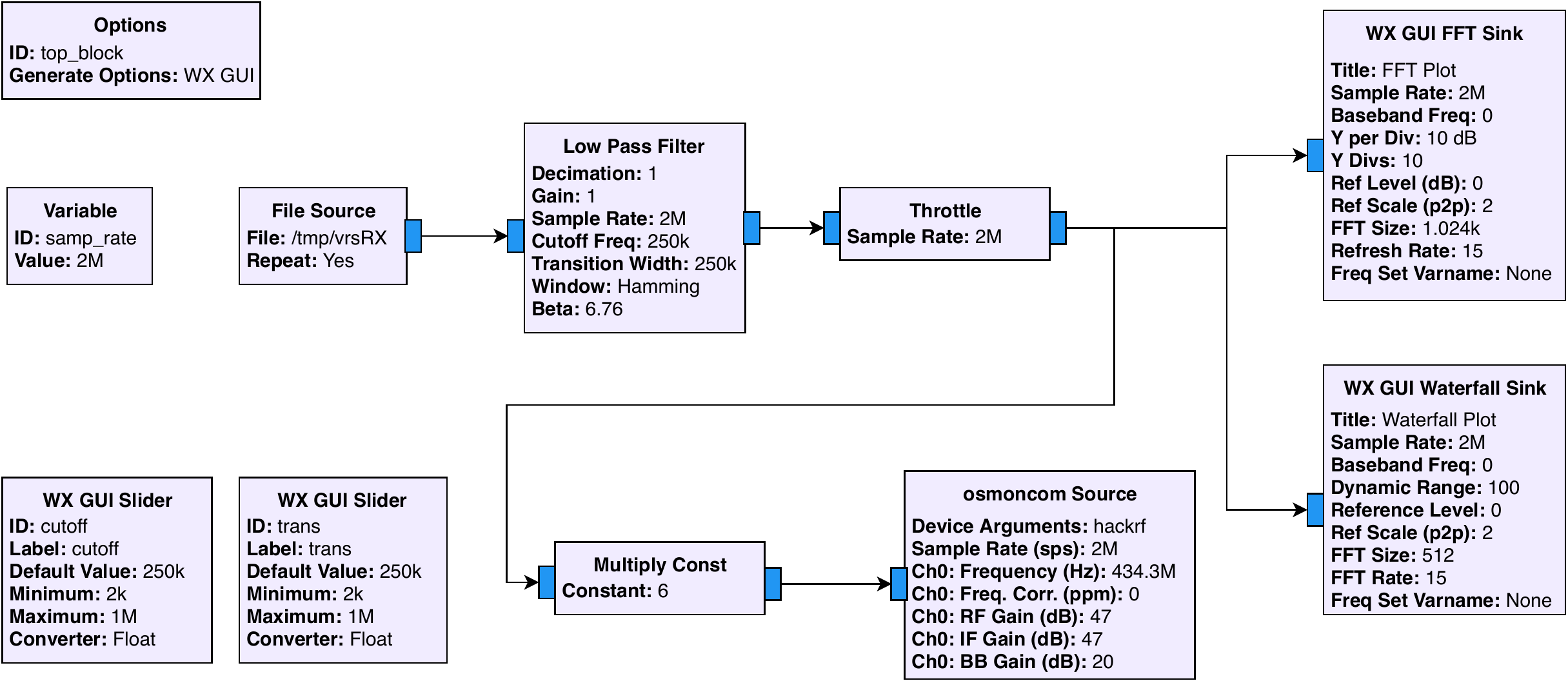}
	\caption{vRSFilter and TX}
	\label{fig:vrsfilterandtx}
\end{figure}

In order to filter down the signal, the captured signal is fed into a Low Pass Filter (LPF) allowing to adjust the cut-off frequency as well as the transition width. This procedure enables to identify the valid signal transmitted by the key fob. Furthermore, the transmission power is increased (i.e. Radio Frequency (RF) and Intermediate Frequency (IF) gains). Allowing to re-transmit the signal to the car and gain full access to the vehicle. 

The method presented allows to by-pass to rolling code of the Skoda vRS key-fob and gain full access to the vehicle. This method can be used against all Skoda range, with the exception of keyless entry which relies on RFID, the method has also been confirmed to work against the Kia Venga, Volvo v40 and Ford Focus, demonstrating the importance to build security-by-design within the automotive industry.

\section{Infotainment System}
\label{sec:info}
\begin{figure}[t]
	\centering
	\includegraphics[width=1\linewidth]{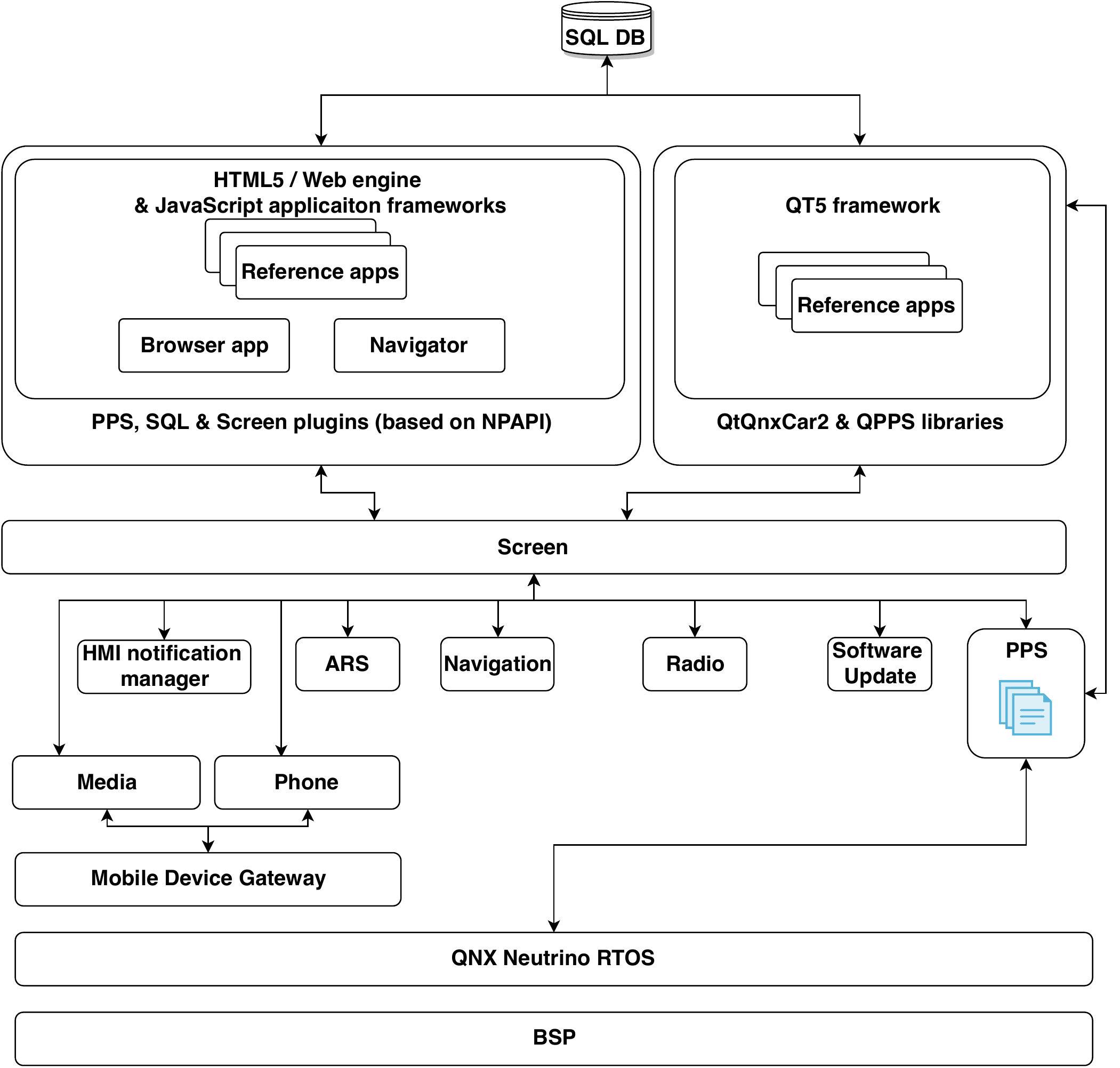}
	\caption{(QNX CAR Platform Infotainment 2.1, 2018) QNX Car Architecture}
	\label{fig:qnx-car}
\end{figure}

The Skoda Octavia vRS offers multiple accessible user interfaces, such as the Modular Infotainment System (MIB). The system was updated for the 2017 Octavia model and features the QNX Operating System (OS) running the Amundsen satellite navigation and online infotainment, additionally also runs Apple Car Play. Figure~\ref{fig:qnx-car} shows the architecture of the operating system.


\begin{figure}[b]
	\centering
	\includegraphics[width=1\linewidth]{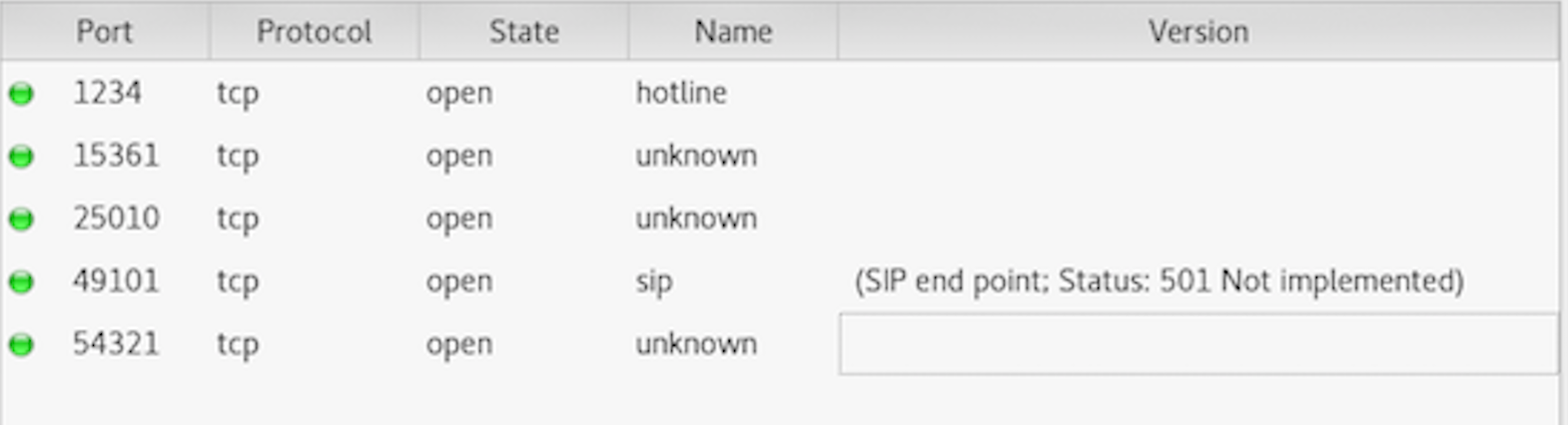}
	\caption{Open Ports}
	\label{fig:ports}
\end{figure}

The QNX OS shown in Figure~\ref{fig:qnx-car} presents a low attack surface. From a red-team perspective, the mobile device gateway, the browser application and the SQL database are the most obvious components. To carry out the reconnaissance phase, the infotainment system was connected to a private WiFi network in order to simulate a compromised connection. The infotainment system was then scanned with Sparta -- a port scanner. Sparta highlighted a number of open ports as shown in Figure~\ref{fig:ports}.

\begin{figure}[t]
	\centering
	\includegraphics[width=1\linewidth]{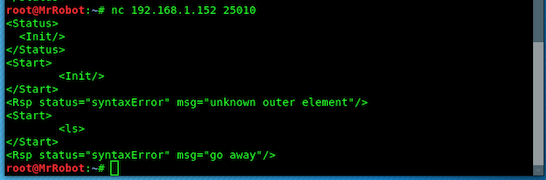}
	\caption{Connection attempt of port 25010}
	\label{fig:25010-nc}
\end{figure}

The open ports were probed for responses. All ports were tested using the NetCat.  Port 25010 responded with XML schema and a "go away message" as shown in Figure~\ref{fig:25010-nc}. While connection to the port was unsuccessful, it is expected that the port is waiting for a specific request in order to return meaningful data.

Amongst the other ports investigated, port 15361 responded to the NetCat requests with information from the infotainment system. The information included the current vehicle speed, GPS location, phone book contact details from the connected iPhone, call history and current mileage-shown in kilometres. This information was provided without authentication requirements. Furthermore, the data was provided in plain text, raising privacy concerns.

\begin{figure*}[tb]
\begin{lstlisting}[language=c, basicstyle=\tiny, label={code:optimpopc}, caption={Information Provided by Port 15361 Upon Connection},captionpos=b]

16:05:02.301 [Info][iMX6.WLAdapter.tsd.wladap] Adapte.   BT{_}APPL{_}PIM{_}DATA{_}IND: parse{_}status=0 (success), 
object{\_}done=1, data(117)=BEGIN:VCARD..VERSION:3.0 ..FN:T*m A**m..N:A**m;T*m..TEL;TYPE=CELL:07765*****7..TEL;
TYPE=HOME:01**4 8****5..END:VCARD

16:05:05.440 [Info][iMX6.Navi.tsd.mibstd2.psd.v15.shortrange.LinkTree]         
VehPos offroad 56.2****4 -3.72****8, 0 cm/s, conf: 1000

16:05:11.579 [Info][iMX6.onlineVHR.OSVHR] [PID=532558 TID= 5] [OSVHR]   CarcomBapKm received from CarCom with current Mileage=11580

16:07:29.817 [Info] [iMX6.WLAdapter.tsd.wladap]  Adapte. BT{_}APPL{_}PIM{_}DATA{_}IND: parse{_}status=0 (success),
object{_}done=1, data(144)=BEGIN:VCARD..VERSION:3.0..FN
:St***rt W**te..N:W**te;St***rt..TEL;TYPE=CELL:0774*****71..X-IRMC-CALL-DATETIME;MISSED:20171106T130835..END:VCARD

16:07:48.191 [Info][iMX6.ASR.asr.engine]  ContextManager:embedGuestContext() Phone.Speech.ContactNames

16:18:56.099[Info][J5e.Radio.System]  Car speed = 0
\end{lstlisting}
\end{figure*}

Listing~\ref{code:optimpopc} provides information of interest, it is, however, important to note that some information has been redacted due to privacy concerns. Additionally to the information displayed in the listing, the information included the address of a distant server, as well as data provided by directly Apple Car Play when an iPhone was connected to the infotainment system including the entire call history i.e. dialled numbers as well as missed calls. It also included the current track which was selected on the iPhone (not playing at the time).  Obtaining this information also suggests that Apple Car Play provides full read access to the phone. 

\begin{figure}[tb]
	\centering
	\includegraphics[width=1\linewidth]{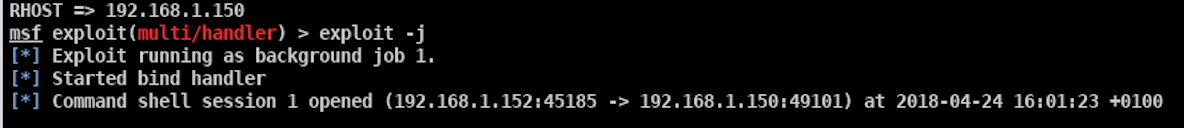}
	\caption{QNX Shell}
	\label{fig:QNX Shell}
\end{figure}

The investigation also highlighted vulnerabilities in the infotainment system. When scanned with Metasploit, a meterpreter command shell was gained as shown in Figure~\ref{fig:QNX Shell} demonstrating the poor security of the system. The meterpreter shell could subsequently be used to execute commands on the infotainment system and potentially create a backdoor or gather more data. 

\subsection{Web Server}
Included in the infotainment system is a Web Browser allowing the user to visit the web and load pre-programmed web-pages. A directory listing was run against the car, the web-server and the web-browser to find hidden directories. 

\begin{itemize}
\item 49101/info
\item 49101/rc
\item 49101/rc/info
\item 192.168.1.150/info
\end{itemize}

\begin{figure}[b]
	\centering
	\includegraphics[width=1\linewidth]{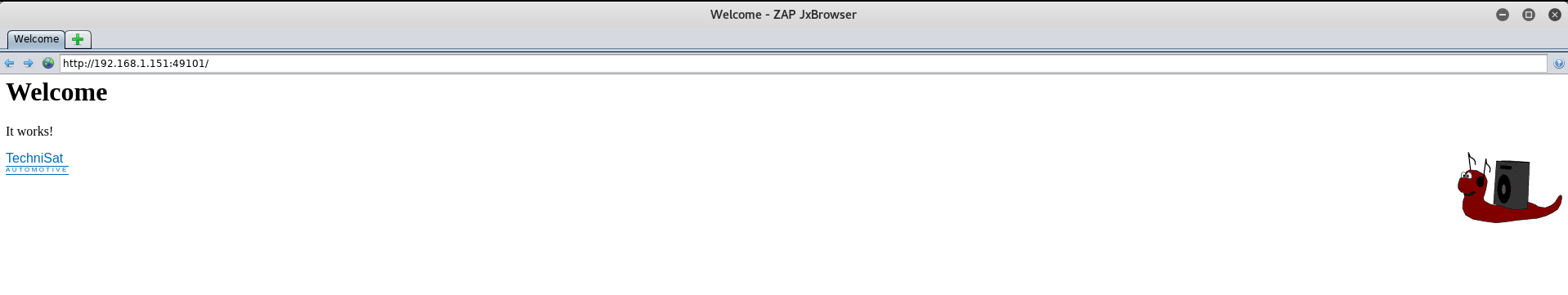}
	\caption{Web Page on Port 49101}
	\label{fig:firefox49101}
\end{figure}

The information obtained through the directory listing was substantial. The QNX OS web-server was found to have discoverable web pages and directories as shown in Figure~\ref{fig:firefox49101}. 

\begin{figure}[tb]
	\centering
	\includegraphics[width=1\linewidth]{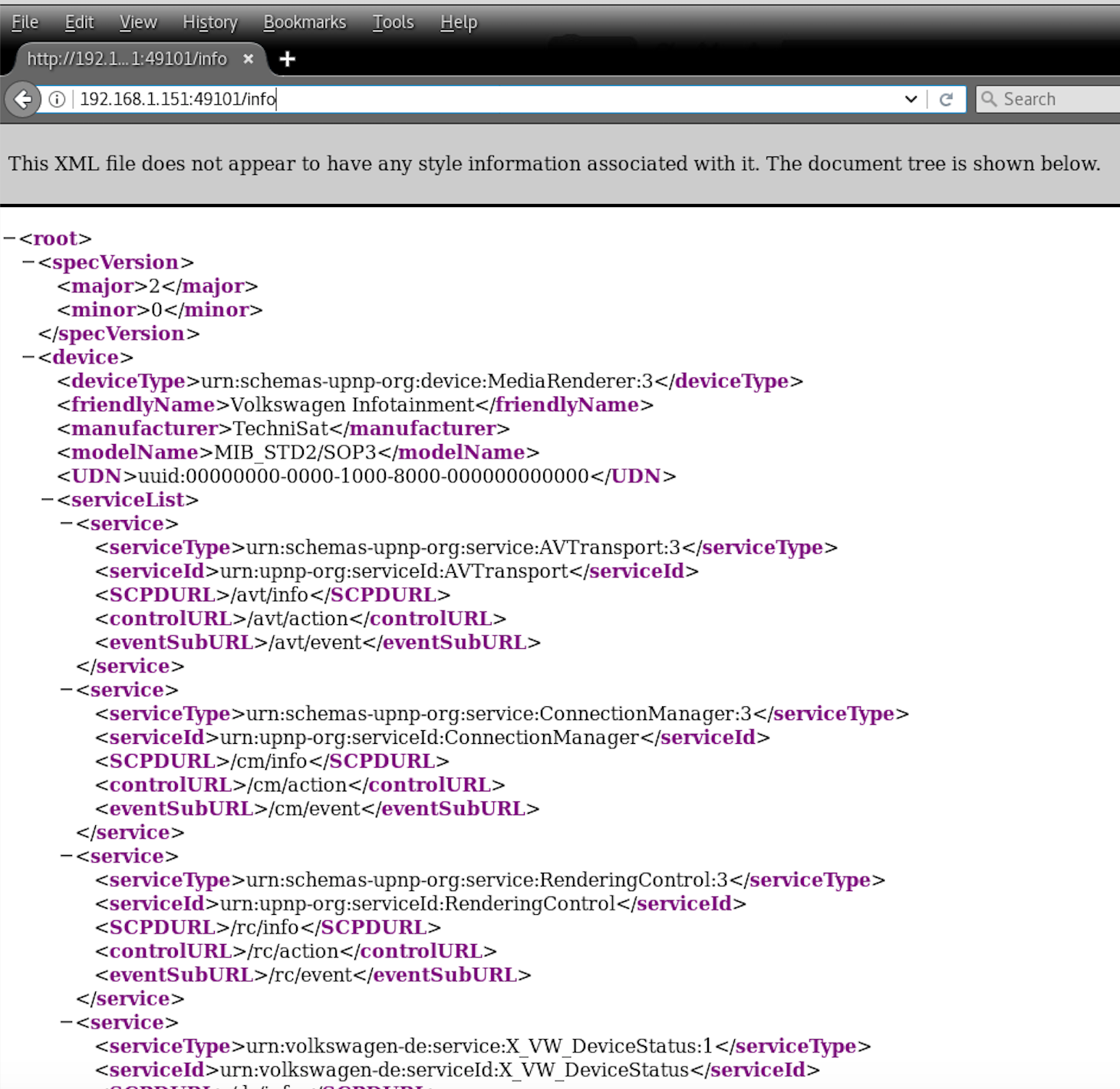}
	\caption{Car Information Provided by Port 49101 under the \textit{/info/} directory}
	\label{fig:49101info}
\end{figure}

\begin{figure}[tb]
	\centering
	\includegraphics[width=1\linewidth]{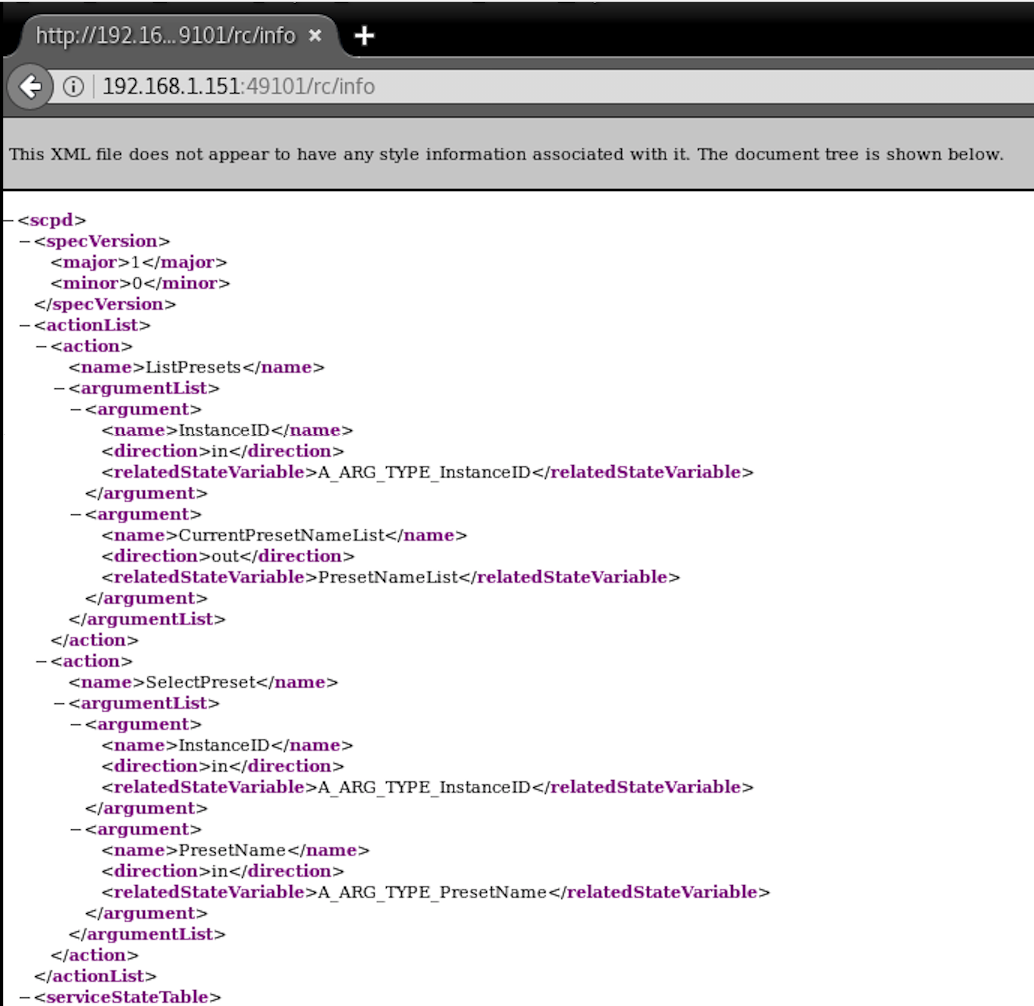}
	\caption{Car Information Provided by Port 49101 Under the \textit{/rc/info} directory}
	\label{fig:rcinfo}
\end{figure}

The following path ``\textit{192.168.1.150/info}", provided an XML Schema, with information relating to sub-directories. As shown in Figure~\ref{fig:49101info} and Figure~\ref{fig:rcinfo} using the information returned from the different path it was possible to browse to specific sections listed in the XML schema. The pages contained plain text information relating to specific request. The information retrieved included GPS Location, Speed, etc, similar to the information gathered in Listing~1.

The infotainment system demonstrated numerous vulnerabilities accessible by an attacker, yielding private and confidential information about the users and the car, raising concerns about end-users privacy. 

\section{On Board Diagnostics Port}
\label{sec:obd}
The On-Board Diagnostics (OBD-II) is part of the accessible user interfaces and has been standard in vehicles since 1996. The OBD-II port is not only a diagnostic interface but also a Controller Area Network (CAN) interface. Typical, CAN packets are 8 bytes in size, however, when necessary larger packets using the Transport Protocol (TP) are being used. The standard TP in use is the ISO-TP (ISO 15762-2), however, vehicles in the Volkswagen and Audi Group (VAG) use their own protocol which is known as VW-TP 2.0 (ISO 14230-3). Additionally, the vRS is fitted with a CAN gateway module which resides prior to the OBD-II port and ensure the exchange of data between the various bus systems. By using the VW-TP the VAG group ensures that no traffic capture is possible through the OBD-II port. Traditionally, a malicious user would connect the USB2CAN or a CANtact to the OBD port to obtain information and capture the traffic.

\begin{figure}[b]
	\centering
	\includegraphics[width=1\linewidth]{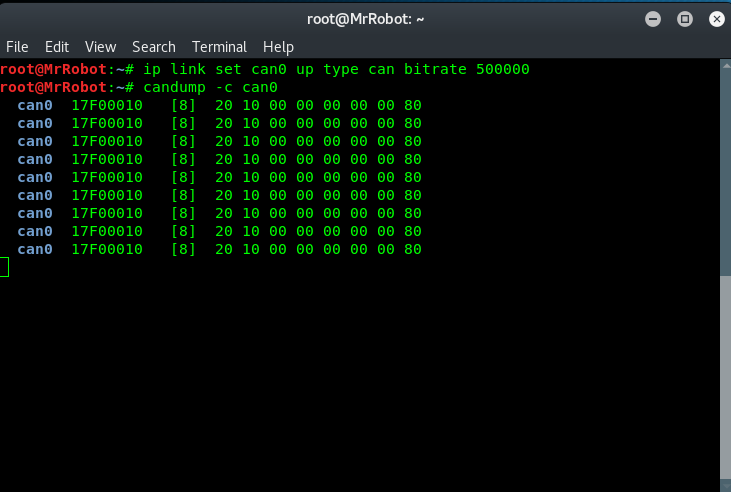}
	\caption{vRS CAN Data; Negative Response Frame}
	\label{fig:vrs-can}
\end{figure}

When trying the same approach on the Skoda Octavia vRS, a negative response is returned by the system as shown in Figure~\ref{fig:vrs-can}. This is due to the VW-TP 2.0 protocol. The protocol requires various parameters to be configured prior to any data being released, crucially the Gateway Module (GM) will only allow specific data requested to flow to the OBD-II port. 

In this section, we present a proof of concept method to gain access to the data of the CAN bus. In order to communicate with it. It is crucial to open a channel, to allow data to be exchanged. Through reverse engineering the communication, it was discovered that this was achieved by sending~\textit{CAN ID 0x200} with the following data~\textit{1f c0 00 10 00 03 01} packet. The command starts the setup with the control unit. Using further reverse engineering techniques, it was possible to highlight the crucial timing to send the command. In fact, without respecting a specific timing in between commands resulted in a connection drop.

\begin{figure}[t]
	\centering
	\includegraphics[width=1\linewidth]{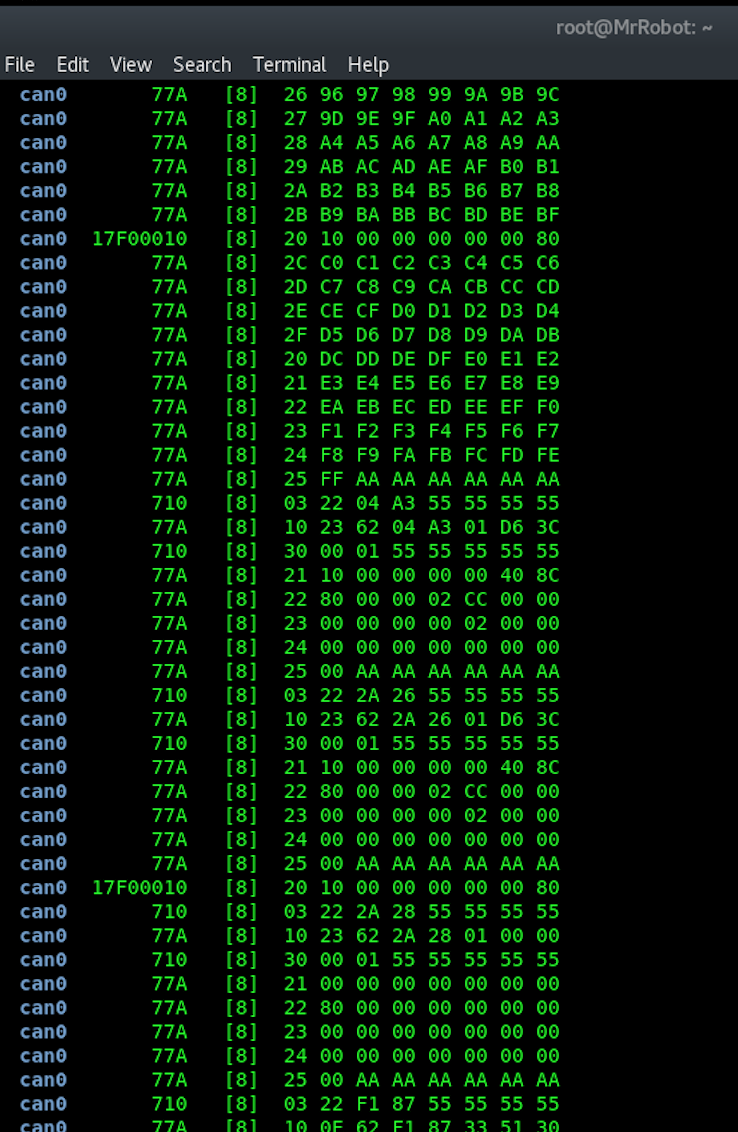}
	\caption{vRS CAN data; Proof-of-Concept Communication with the CAN Bus}
	\label{fig:vrs-module-dump}
\end{figure}

\begin{figure}[b]
	\centering
	\includegraphics[width=1\linewidth]{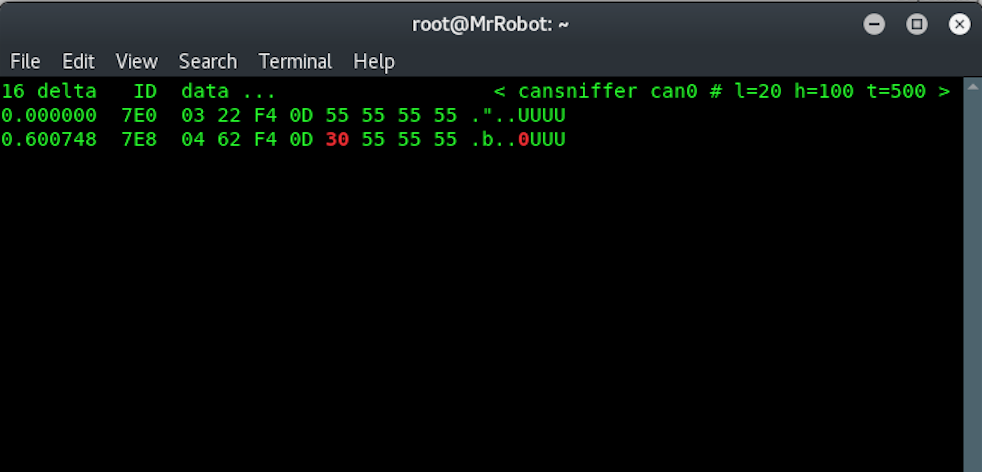}
	\caption{vRS Speed}
	\label{fig:vrs-speed}
\end{figure}

Figure~\ref{fig:vrs-module-dump} shows the CAN data obtained on the Skoda vRS. The proof of concept enables to capture live data on the bus and obtain data such as the engine speed. Figure~\ref{fig:vrs-speed} provides the speed of the car whilst in motion (in red) recorded at 48 miles per hour. 

While this method enables to capture live data from the system such as diagnostic information, the proof of concept also enables to modify parameters and inject commands. There is a number of tools provided by VAG available for garage mechanics to diagnose faults by accessing the OBD-II port and collecting CAN data, two of these tools are the Snap-On SOLUS Scan Tool and the Ross-Tech VCDS. Both authenticate with the gateway module/ECU and are able to retrieve data and modify parameters. While both tools are publicly available and can be used to change parameters in the vehicle, both have prices ranging from \pounds200 to \pounds350. The algorithm provided within this manuscript provides similar access to the CAN bus. 

\begin{figure}[tb]
	\centering
	\includegraphics[width=1\linewidth]{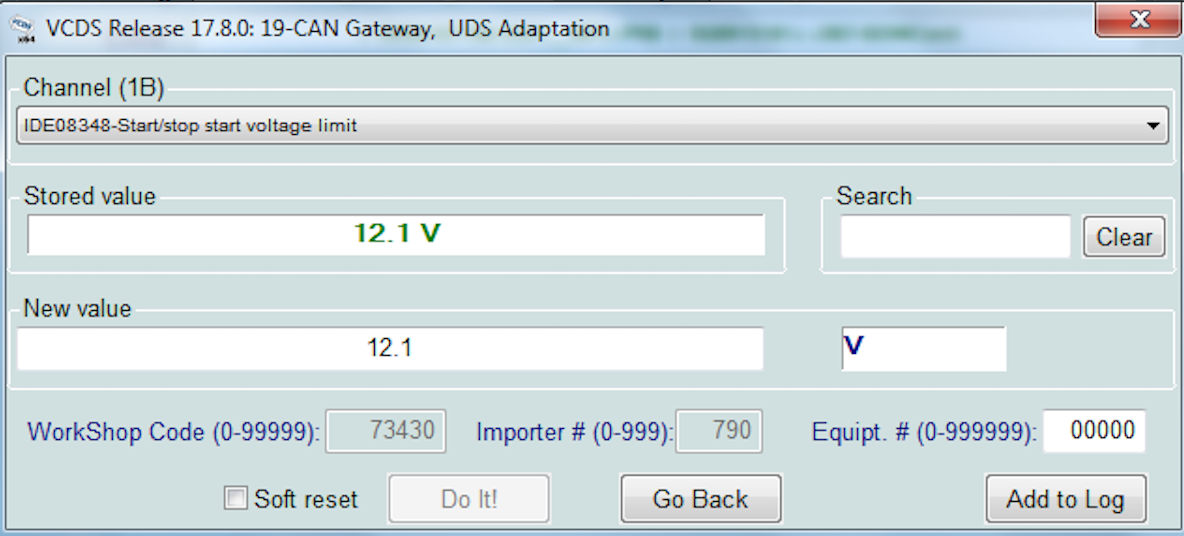}
	\caption{VCDS Voltage limit adjusted}
	\label{fig:voltage-limit}
\end{figure}

Furthermore, the CAN bus accessible through the OBD-II port enables to modify central components of the car such as the automatic start-stop feature. Using the CAN Gateway and selecting adaptation,  the automatic start-stop feature was disabled, without displaying any error message to the user. Disabling this feature requires to modify the ``IDE08348-Start/stop voltage limit" from 7.5v to 12.1v resulting in a failure of the system to operate. Similarly, this can be achieved using the VCDS system as shown in Figure~\ref{fig:voltage-limit}.

As the voltage is below the 12.1V level when the car is started the automatic start/stop is permanently disabled with no error message visible to the user. The lack of authentication on the OBD-II port can lead to malicious users modifying the behaviour of the car. For example, the OBDEleven device, a snap-on device which connects to the OBD-II port and allows for full system diagnostics, as well as monitoring and the ability to change parameters. The OBDEleven differs from devices through its small form factor. When connected, the device is invisible to the driver and works with Bluetooth in conjunction with an Android application. 

The OBDEleven device was also used within this manuscript alongside a LG Nexus 5 to demonstrate its efficacy as a malicious device and highlight the strong requirements for authentication on OBD-II ports. The OBDEleven allows for example to activate the reverse mirror dip with a single click. Furthermore, it is also possible to enable and configure "high beam assist" on the Skoda vRS, which if ordered from Skoda at a cost \pounds315. While these examples are trivial, the device was also capable of configuring the infotainment system to allow access to the engineering (green) menu where it was possible to view configurations and modify the debugging destinations which are not visible to the public. To access this menu the following procedure was followed: I) Once connected, the ``Control unit" was selected and changed to ``Change Service". II) ``Developer Mode" was then selected from the options. III) Once complete the root menu was then accessed and ``adaptions" selected and ``Developer Mode" was enabled and applied. IV) After the configuration was complete on the OBDEleven the menu button on the infotainment system was held in for 5 seconds providing access to the ``Service mode" menu. V) from there ``Testmode" and ``Green Engineering Menu" were selected allowing access to the Engineering menu.

\begin{figure}[h]
	\centering
	\includegraphics[width=1\linewidth]{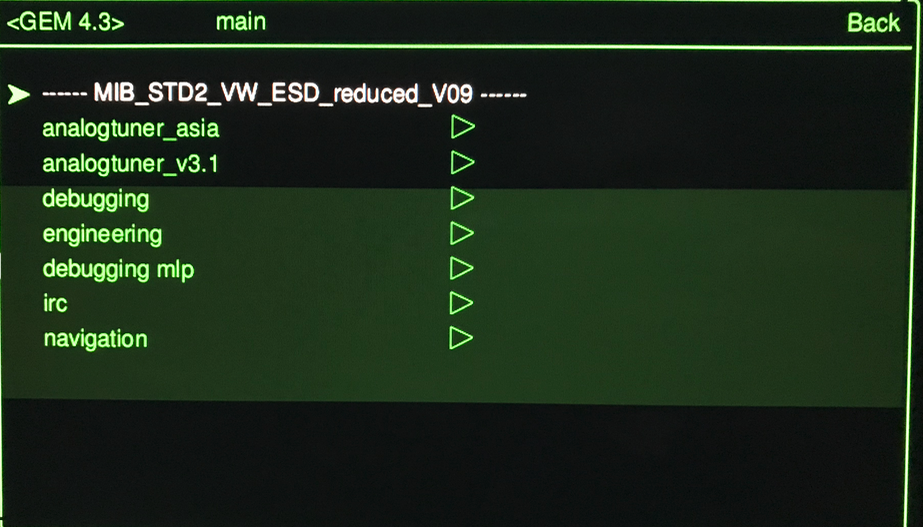}
	\caption{Engineering Menu}
	\label{fig:engineering-menu}
\end{figure}

\begin{figure}[b]
	\centering
	\includegraphics[width=1\linewidth]{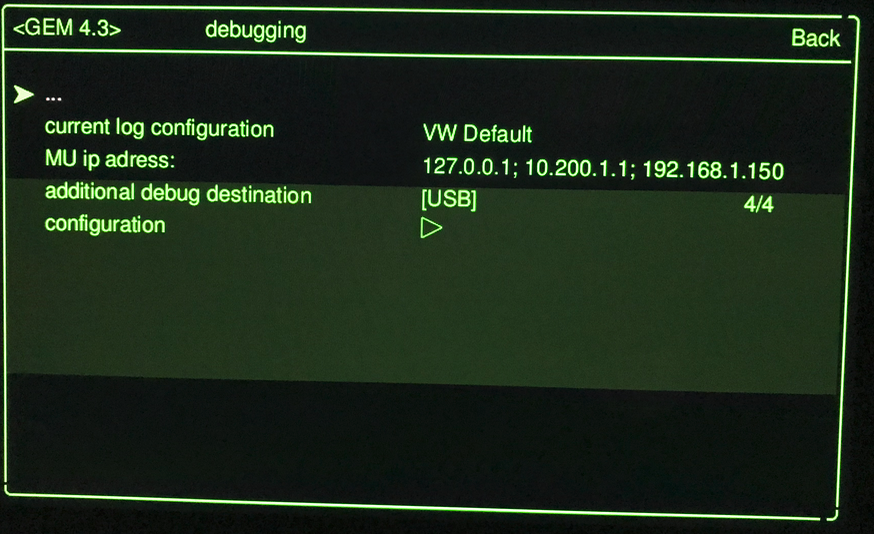}
	\caption{Debugging Menu}
	\label{fig:debugging-menu}
\end{figure}

Figure~\ref{fig:engineering-menu} and Figure~\ref{fig:debugging-menu} highlight options to change the debugging destination to a USB drive, including logs of the vRS. Additionally, it is possible to configure the menu to provide networking logs, phone logs and the VW Default logs, including information about the state of the car and its connection status with VAG further highlighting security and privacy concerns. 

\begin{table*}[tb]
    \centering
    \resizebox{\textwidth}{!}{
    \begin{tabular}{|c | c c c c c c c c c c c c c c c c c c c c c}
        \hline
        & \multicolumn{21}{|c}{Fuel Quantity (RPM, EGT)/mg/cyc}\\
        1/min & 0 & & 551 & & 1250 & & 1750 & & 2016 & & 2499 & & 3000 & & 3500 & & 4000 & & 4500 & & 5200 \\
        & & 500 && 1000 && 1500 && 1900 && 2247 && 2750 &&  3250 && 3750 && 4250 && 4800 & \\
        \hline
        \rowcolor{gray_2}
        \cellcolor[gray]{1}500 & \cellcolor[gray]{1}30.00 & \cellcolor[gray]{1}30.00 & \cellcolor[gray]{1}30.00 & \cellcolor[gray]{1}30.00 & \cellcolor[gray]{1}39.60 & 48.00 & 54.00 & 56.00 & 54.75 & 53.50 & 52.50 & 51.75 & 51.00 & 50.00 & 49.50 & 48.00 & 47.00 & 44.00 & 39.00 & 32.00 & \cellcolor[gray]{1}0.00 \\
        
        \rowcolor{gray_2}
        \cellcolor[gray]{1}900 & \cellcolor[gray]{1}30.00 & \cellcolor[gray]{1}30.00 & \cellcolor[gray]{1}30.00 & \cellcolor[gray]{1}30.00 & \cellcolor[gray]{1}39.60 & 48.00 & 54.00 & 56.00 & 54.75 & 53.50 & 52.50 & 51.75 & 51.00 & 50.00 & 49.50 & 48.00 & 47.00 & 44.00 & 39.00 & 32.00 & \cellcolor[gray]{1}0.00 \\
        
        \rowcolor{gray_2}
        \cellcolor[gray]{1}1000 & \cellcolor[gray]{1}30.00 & \cellcolor[gray]{1}30.00 & \cellcolor[gray]{1}30.00 & \cellcolor[gray]{1}30.00 & \cellcolor[gray]{1}39.60 & 48.00 & 54.00 & 56.00 & 54.75 & 53.50 & 52.50 & 51.75 & 51.00 & 50.00 & 49.50 & 48.00 & 47.00 & 44.00 & 39.00 & 32.00 & \cellcolor[gray]{1}0.00 \\
    \end{tabular}
    }
    \caption{Power Delivery Text Based Interface}
    \label{tab:my_label}
\end{table*}

\section{Electronic Control Unit}
\label{sec:ECU}
The Electronic Control Unit (ECU) is controlling vital aspects of the vehicle including the fuel to air mix, turbo boost pressure, etc\ldots. The ECU is another accessible user interface of the vehicle. Over the years, different tuning of the ECU has appeared. ECU tuning is commonly referred to as Chip tuning - Chipping was the process of physically removing the ECU from the vehicle, open the ECU in a controlled environment, using a K-tag tool to connect to it, as well as configure the pin out/probes to connect to specific connection locations within the ECU. While this method is effective, it is time consuming and requires physical access to the ECU. Additionally, this method also presents a greater risk of damaging the ECU. A more practical option is to tune the ECU through the OBD-II port, however, as the Skoda vRS employs the VW-TP 2.0 protocol, modifying the ECU is not as straight forward as on other vehicles. 

In this section, the process of chip tuning is described, furthermore, we highlight the risks and possible adverse effects chip tuning may have on the vehicle if achieved by a malicious user. In this section, we also demonstrate how chip tuning can be achieved using off-the-shelf tools.

The most common chip tuning application, is the process of changing the driving parameters, also referred to, as ``remapping". Remapping consists of taking a reading of the processing chip within the ECU which stores the engine map, contained within the map are details on boost pressure, ignition, throttle position, fuel pressure, etc\ldots.  


Remapping the ECU can be achieved through readily available tools such as \textit{CMDFlash OBD} and \textit{WinOLS}. CMDFlash is a flash programmer which supports various protocols including EOBD K-Line, as well as, EOBD CAN Bus amongst others. WinOLS is an application for Windows which is capable of searching for engine maps within the ECU. The tool is able to successfully modify parameters in order to change the map/behaviour of the vehicle. It is possible to modify parameters by entering specific values or by modifying a 3D map/visual representation of the engine power delivery.

To change the map of the engine it is necessary to connect the CMDFlash tool to the OBD-II port, once the CMDFlash identified the model of vehicle as the vRS the ECU data is downloaded and saved. The ECU data can then further be used by WinOLS. 

WinOLS provides a range of modification options, ranging from graphical interface to a visual modifier. Table~\ref{tab:my_label} highlights the text-based interface. In this section, we modified the torque limiter, both through the 3D model and through the text-based interface.  

\begin{figure}[tb]
	\centering
	\includegraphics[width=0.7\linewidth]{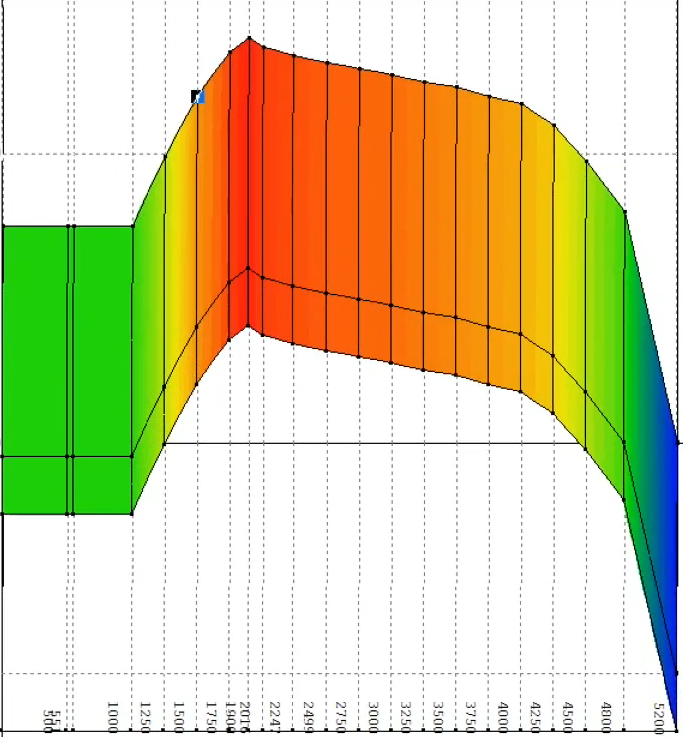}
	\caption{Original Power Delivery}
	\label{fig:map-of-original-delivery}
\end{figure}

The 3D model available provides a visual representation of the current power delivery method in the engine. The original power is shown in Figure~\ref{fig:map-of-original-delivery}.

\begin{figure}[tb]
	\centering
	\includegraphics[width=0.8\linewidth]{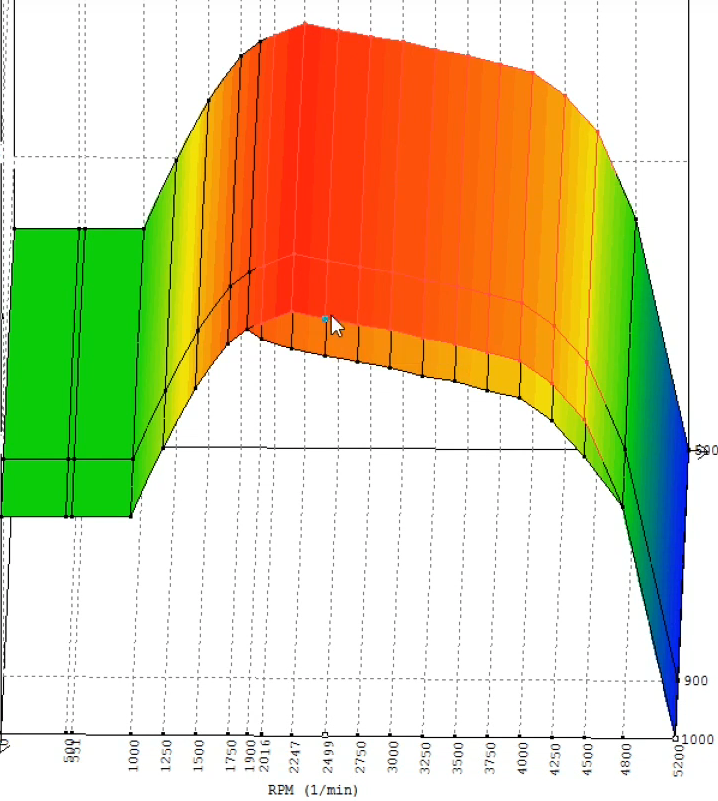}
	\caption{Modified Power Delivery}
	\label{fig:modified-power-delivery}
\end{figure}

As shown in Figure~\ref{fig:map-of-original-delivery} the power delivery peaks at over 1900 Revolutions Per Minute (RPM), the torque then reduced until 3250rpm where the reduction becomes much more pronounced (i.e. 4000rpm) the performance drop-off forces a gear change due to the degradation. Using the 3D model, the area post 1900rpm until 4500rpm can be highlighted allowing for the selected area to be modified  (increased or decreased), affecting the overall performance of the vehicle as shown in Figure~\ref{fig:modified-power-delivery}.

Adjusting the power delivery post 1900rpm creates a much more usable and evenly distributed powerband which changes the driving characteristics of the vehicle. Additionally, the 3D model also allows visualising the comparison between the original parameters and the modified ones as shown in Figure~\ref{fig:modified-power-delivery}.

While remapping the engine is often used to improve the drivability of the vehicle, an infected computer, or and infected tool could lead to detrimental effects both to the engine and possibly the user.  

\begin{figure}[b]
	\centering
	\includegraphics[width=1\linewidth]{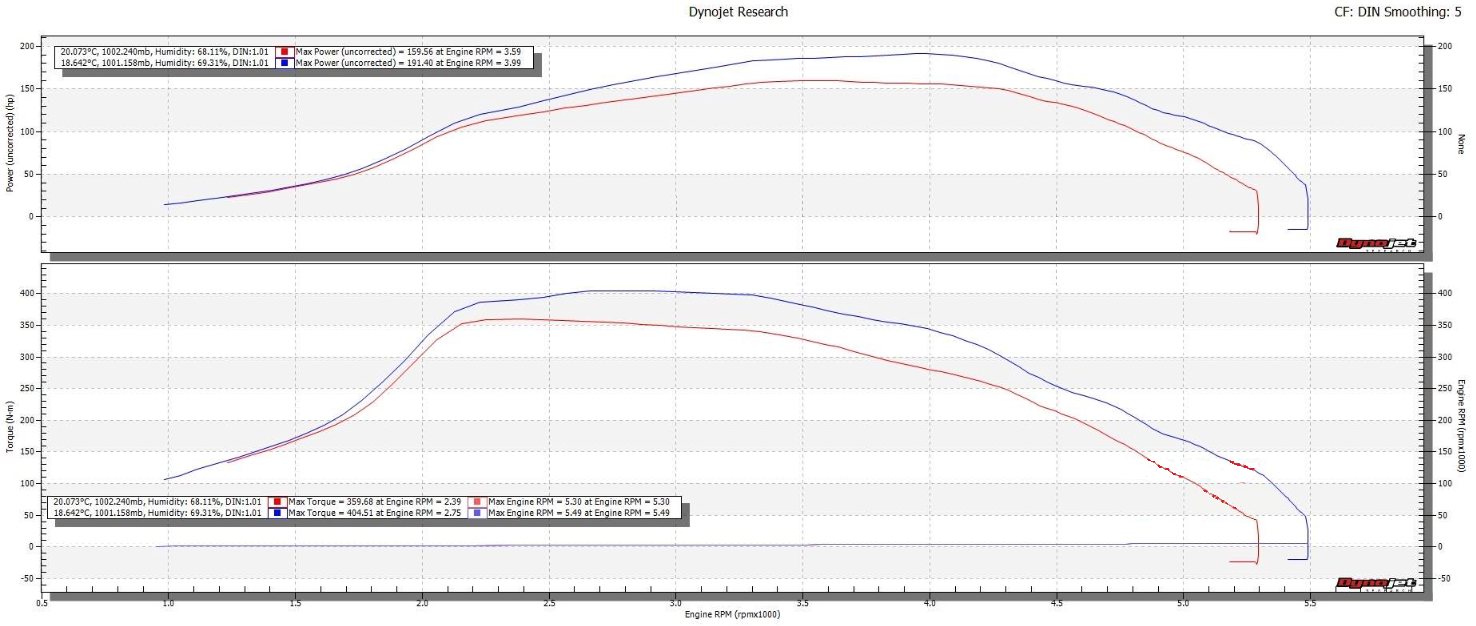}
	\caption{Remap Result Comparison}
	\label{fig:comparison}
\end{figure}

Once modified, the changes were uploaded to the ECU through the OBD port. The car was then placed on a "rolling road" to measure the performances. The engine alteration is displayed in Figure~\ref{fig:comparison}. The results represent the power at the wheels, rather than at the flywheels (e.g. The car demonstrates 184bhp at the flywheel, however only 159.56 Bhp at the wheels).

\begin{figure}[tb]
	\centering
	\includegraphics[width=1\linewidth]{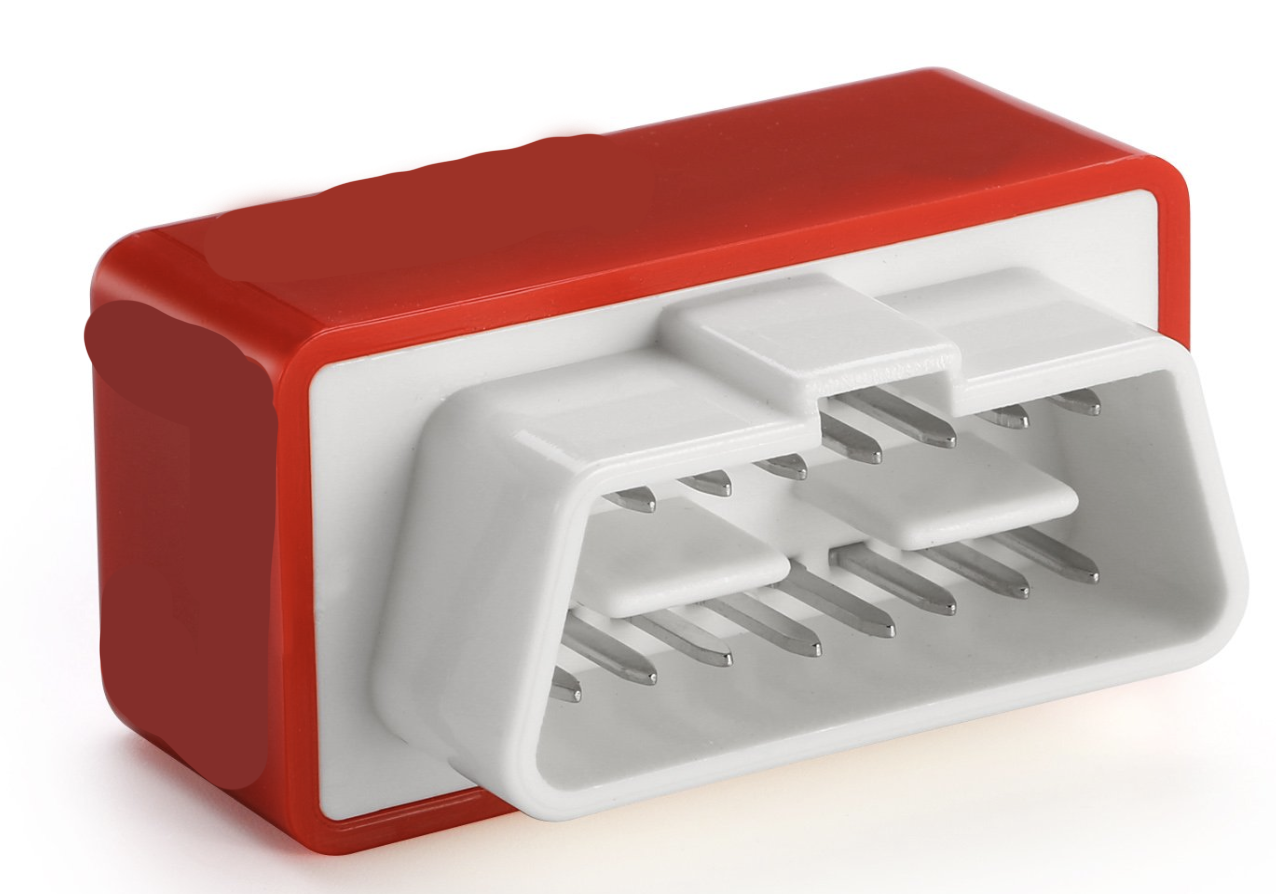}
	\caption{OBD plug and play device increasing the car performance}
	\label{fig:CAN_HACK}
\end{figure}

The modification of the ECU lead to a recorded gain of 32 break horsepower at the wheels and a torque increase of 45Nm, furthermore, the remap also lead to an additional 7mpg savings. While this experiment demonstrates the benefits of remapping, a malicious script can easily alter the performances of the car during remapping and cause serious damage to the car, the engines and the occupants. Furthermore, it is now possible to find "plug-and-play" devices claiming to increase the performance of the car as shown in Figure~\ref{fig:CAN_HACK}.

\begin{figure*}[tb]
	\centering
	\includegraphics[width=1\linewidth]{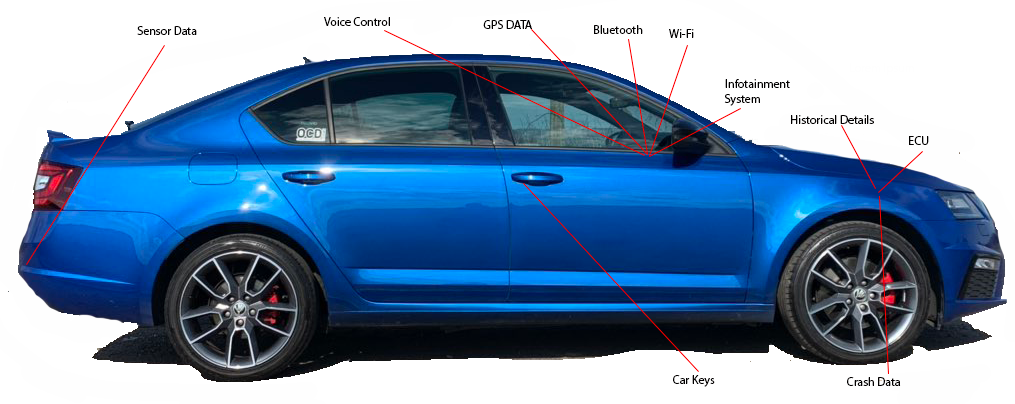}
	\caption{Skoda Octavia vRS Sensors}
	\label{fig:Skoda}
\end{figure*}

Throughout our study, half a dozen plug-and-play devices sold by unknown sellers have been identified, while these may well increase the performance of the car, they could also dramatically deteriorate the car performances and be used to infect the car with a malware, or into performing an unwanted action. It is therefore essential to be able to identify components that may have been tempered with. 

\section{Vehicular Forensic Components}
\label{sec:forensics}
In light of the different vulnerabilities and Vehicular Security Testing methodology proposed in this manuscript, numerous components holding key information about the state of the car have been identified. This section highlights components of the vehicles, containing data that can be used during a digital forensic investigation. 

Figure~\ref{fig:Skoda} lists a number of the components holding key data digital forensic investigators can use to investigate a crime within the automotive industry. Furthermore, data can be located using the Car Security Testing Methodology presented in Figure~\ref{fig:sectTesting}, starting with the identification of the vehicular technology under investigation. An Informative list of components is presented hereafter to facilitate the location of data in vehicular technologies.

\subsection{Infotainment Forensics}
Modern infotainment systems contain numerous information about the users including Bluetooth ID from paired phones and phones that may have tried to connect. Information such as call logs, can also be found, even when Android Auto or Apple Car play are not enabled. The Infotainment system also contains information on the music files 
Pictures such as thumbnails that can have been accessed through the infotainment system. Voice data can also be recovered on some models
when the ``Voice Control" commands have been activated. This information can be crossed-examined during a forensic investigation to provide historic details on the usage of the vehicle. 

\subsection{GPS Forensics}
Global Positioning System (GPS) and Glonass and Galileo data can be used to find the last location driven, including last destinations and stops made by the vehicle in locations it stopped.

\subsection{ECU Forensics}
By removing the ECU or accessing data through the OBD-II port, forensics investigators can retrieve information such as the mileage readings, the Vehicle Identification Number (VIN) \& Serial Numbers. This information can be used to formally identify the car. Furthermore, a History of faults, the time they were triggered, and their mileage at the time of trigger. Some vehicles also allow for crash data to be recovered alongside a history of resolved faults and events, such as the brakes information (i.e. where the brakes applied, what light was turned on, etc\ldots). The ECU also contains external memory, than can be interfaced with $I^2C$ or SPI, to obtain information. The ECU might also contain a JTAG interface, or a UART that might be accessible. 

\subsection{Key fobs Forensics}
Modern key fobs contain numerous information about the car such as the VIN number, the transponder ID, or the number of keys associated with the vehicle and the ID of the key.  Some vehicles also store the last mileage readings, the fuel status of the car and vehicle data such as historical faults, and crash data. 
\newline 

The information provided by the different components of the car during a digital forensic investigation can help answer key question after a crash or a crime and help digital forensic investigators to uncover  the speed of the car, the angle of the steering wheel, the time and impact (i.e. from the back or the front) or information about the Automatic Collision Notification (ACN) system.

\section{Conclusion}
\label{sec:conclusion}
Vehicular technologies are evolving rapidly with embedded connection and self-driving modules. While the range of services increases for the end-user, these modules drastically increase the attack surface. There has been little documented research on the cyber-security aspects of vehicular technologies currently on the road. In this manuscript vehicular security testing methodology was presented, enabling cyber-security researchers to identify key components of vehicles. This methodology was subsequently used to evaluate the cyber-security of a Skoda Octavia vRS. The key fob was demonstrated to be vulnerable to rolling code by-pass attacks enabling a malicious user to get access to the vehicle. Furthermore, it was demonstrated that when accessed the infotainment system was yielding private user information, that the system was vulnerable to exploit, that information relating to the speed and location of the vehicle was also exposed. In this manuscript, the OBD port was also investigated. The VW TP 2.0 protocol was reversed engineered, enabling a malicious user to gather information about the system, such as the speed of the vehicle, furthermore, the proof-of-concept also enabled to modify key elements of the vehicle such as the start-stop feature. Moreover, the ECU was investigated through the OBD-II port. It was demonstrated that the behaviour of the car could be altered and the danger of off-the-shelf OBD-II ECU remapping devices was highlighted. Finally, key components of the car containing key data about the vehicle were identified for digital forensic investigators to investigate.  

While the security investigation focused primarily on the Skoda Octavia vRS 2017, the methodology provided is generic and applied to all types of vehicles. Furthermore, all elements and security weaknesses identified, may be replicated on other vehicles. Finally, the components identified for digital forensic investigation described in this manuscript are present in all modern vehicles. It is, therefore, crucial to improve the security of vehicular technologies to ensure the physical security of the occupants, while essential to provide digital forensics investigator with data to investigate crash information or crimes committed using vehicular technologies. Future work should focus on the security and digital forensic investigation of vehicles compliant with the latest UNECE WP.29 standards, the virtualisation of ECU components as well as V2X communications. 

\section{Acknowledgements}
All the vulnerabilities disclosed within this paper followed a responsible disclosure procedure. We would like to particularly thank \v{S}koda Auto A.S teams for their prompt reply to the vulnerabilities highlighted within this manuscript and their continuous work towards improving the cyber-security of vehicular technologies and user safety. 

\bibliographystyle{IEEEtran}
\bibliography{mybibfile}

\begin{IEEEbiography}[{\includegraphics[width=1in,height=1.25in,clip,keepaspectratio]{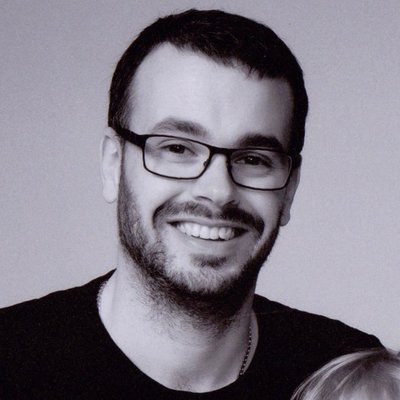}}]{Colin Urquhart} 
 received a Bachelor with Honours in Ethical Hacking and Computer Security from Abertay University in Dundee, Scotland. Colin is currently looking for a Ph.D. in vehicular security or vehicular digital forensics. His research interests include security, malware analysis, reverse engineering and digital forensics for cyber-physical systems.
\end{IEEEbiography}

\begin{IEEEbiography}[{\includegraphics[width=1in,height=1.25in,clip,keepaspectratio]{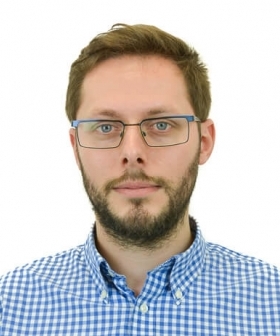}}]{Xavier Bellekens} 
received the Bachelor Degree from Henam in Belgium; the Masters degree in Ethical Hacking and Computer Security from the University of Abertay Dundee and the Ph.D. in Electronic and Electrical Engineering from the University of Strathclyde in Glasgow in 2010, 2012 and 2016 respectively.He is currently a Chancellor's Fellow Lecturer in the department of Electronic and Electrical Engineering at the University of Strathclyde working on cyber-security for critical infrastructures. Previously, Xavier was a Lecturer in Security and Privacy at the University of Abertay in Dundee within the Department of Cyber-Security where he lead the Machine Learning for Cyber-Security research group. His current research interests include machine learning for cyber-security, autonomous distributed networks, the Internet of Things and critical infrastructure protection.
\end{IEEEbiography}

\begin{IEEEbiography}[{\includegraphics[width=1in,height=1.25in,clip,keepaspectratio]{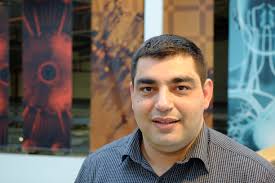}}]{Christos Tachtatzis} 
is a Senior Lecturer Chancellors Fellow in Sensor Systems and Asset Management, at the University of Strathclyde. He holds a BEng (Hons) in Communication Systems Engineering from University of Portsmouth in 2001, an MSc in Communications, Control and Digital Signal Processing (2002) and a PhD in Electronic and Electrical Engineering (2008), both from Strathclyde University. Christos has 12 years experience, in Sensor Systems ranging from electronic devices, networking, communications and signal processing. His current research interests lie in extracting actionable information from data using machine learning and artificial intelligence.
\end{IEEEbiography}

\begin{IEEEbiography}[{\includegraphics[width=1in,height=1.25in,clip,keepaspectratio]{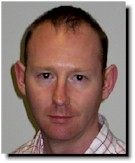}}]{Robert Atkinson} 
received the B.Eng. (Hons.) degree in electronic and electrical engineering; the M.Sc. degree in communications, control, and digital signal processing; and the Ph.D. degree in mobile communications systems from the University of Strathclyde, Glasgow, U.K., in 1993, 1995, and 2003, respectively. He is currently a Senior Lecturer at the institution. His research interests include data engineering and the application of machine learning algorithms to industrial problems including cyber-security.
\end{IEEEbiography}

\begin{IEEEbiography}[{\includegraphics[width=1in,height=1.25in,clip,keepaspectratio]{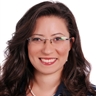}}]{Hanan Hindy} 
is a second year Ph.D. student at the Division of Cyber-Security at Abertay University, Dundee, Scotland. Hanan received her bachelor degree with honours (2012) and a masters (2016) degrees in Computer Science from the Faculty of Computer and Information Sciences at Ain Shams University, Cairo, Egypt. Her research interests include Machine Learning and Cyber Security. She is currently working on utilising deep learning for IDS.
\end{IEEEbiography}

\begin{IEEEbiography}[{\includegraphics[width=1in,height=1.25in,clip,keepaspectratio]{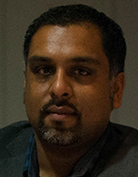}}]{Amar Seeam} 
is a Senior Lecturer in Computer Science at Middlesex University Mauritius. He obtained his PhD from the University of Edinburgh in 2015. Other qualifications obtained by Amar include a BEng (Hons) in Mechanical Engineering (2003) and MSc in Information Technology (2004) conferred by the University of Glasgow as well as an MSc in System Level Integration from the University of Edinburgh (2005). His research interests include Simulation Assisted Control, Cybersecurity, Internet of Things and Building Information Modeling.
\end{IEEEbiography}

\EOD

\end{document}